\documentclass[12pt]{article}
\usepackage{geometry}
\usepackage{setspace}
\usepackage{caption}
\usepackage{times}
\usepackage{graphicx} 
\usepackage{wrapfig}
\usepackage{dcolumn}
\usepackage{bm}
\usepackage{subfigure}
\usepackage{amssymb,amsmath}
\usepackage{hyperref}
\usepackage[super,sort&compress]{natbib}

\title{Dynamic-Kinetic Duality \\of Particulate and Multiphase Systems}
\author{Carlos E. Colosqui\\
Department of Mechanical Engineering, Stony Brook University\\
Department of Applied Mathematics \& Statistics, Stony Brook University\\
\texttt{carlos.colosqui@stonybrook.edu}}
\date{}

\begin{document}
\maketitle

\begin{abstract}
The evolution of particulate and multiphase systems can transition from dynamic regimes, governed by classical transport equations with well-defined damping coefficients, to anomalously slow relaxation described by rate equations when the system is critically close to equilibrium. 
This regime crossover has been both theoretically predicted and experimentally observed in diverse multiphase systems relevant to numerous technological applications, including nanoparticle adhesion at interfaces and liquid imbibition under microscale confinement, and it is attributed to the presence of small nanoscale features of physical or chemical nature at liquid-solid interfaces. 
This article presents a theoretical framework to more accurately predict and control, advancing or delaying, the dynamic-to-kinetic regime crossover, and highlights strategies for harnessing this phenomenon to enhance or suppress different transport processes in confined multiphase systems. 
\end{abstract}


\section{Introduction}
Particulate and multiphase systems are conventionally modeled as consisting of homogeneous phases separated by sharp, ideally smooth interfaces, such as perfectly flat or spherical boundaries. This description yields mechanical or thermodynamic potentials \( U = U(\mathbf{q}) \) as a function of a set of observable state variables \( \mathbf{q} \) that exhibit a single, well-defined global minimum. This minimum defines the equilibrium state for a unique set of observables \( \mathbf{q}_E \). These idealizations, which align with the framework of Gibbs sharp-interface thermodynamics,\cite{gibbs1906scientific,rowlinsonmolecular} enable the formulation of mathematically tractable models that predict system dynamics and time evolution.
However, this conventional description overlooks a key physical phenomenon: liquid-fluid and solid-fluid interfaces exhibit nanoscale physical and chemical features that introduce energetic perturbations \( \Delta U \gtrsim k_B T \), which are comparable to or exceed the thermal energy of the system (where \( k_B \) is the Boltzmann constant and \( T \) is the absolute temperature).\cite{colosqui2013,colosqui2017}

The neglect of such fine-scale energy perturbations is often appropriate when sufficiently far from equilibrium, but it severely limits the ability to predict and rationalize near-equilibrium behaviors. It is well established that surfaces with sufficiently large topographic feature dimensions induce observable (coarse-scale) metastable equilibrium configurations, commonly identified as the coexisting Cassie and Wenzel states in the case of liquid wetting a ``rough'' solid surface.\cite{cassie1944wettability,wenzel1949surface,bico2001rough}
It is also known that ``rough'' energy landscapes with a multiplicity of metastable wetting states arise for complex surface topographies, and substantial energy barriers can prevent transitions between coarse-scale (observable) metastable states.\cite{whyman2011,giacomello2012,mihalis2012,mihalis2013,bormashenko2015progress,al2022toward}
Similarly, fine-scale energy fluctuations arising over small spatial scales \( \ell \), which are often not accessible by experimental techniques, generate rough energy landscapes with densely packed metastable states that are not directly observable.\cite{zwanzig1988}
As a result, the observable behavior of near-equilibrium systems can exhibit nontrivial time evolutions that cannot be adequately described using conventional diffusivity constants or regime-independent transport coefficients.\cite{zwanzig1988,banerjee2014diffusion,colosqui2019diffusion}

\begin{figure}[h!]
\begin{center}
\includegraphics[width=\textwidth]{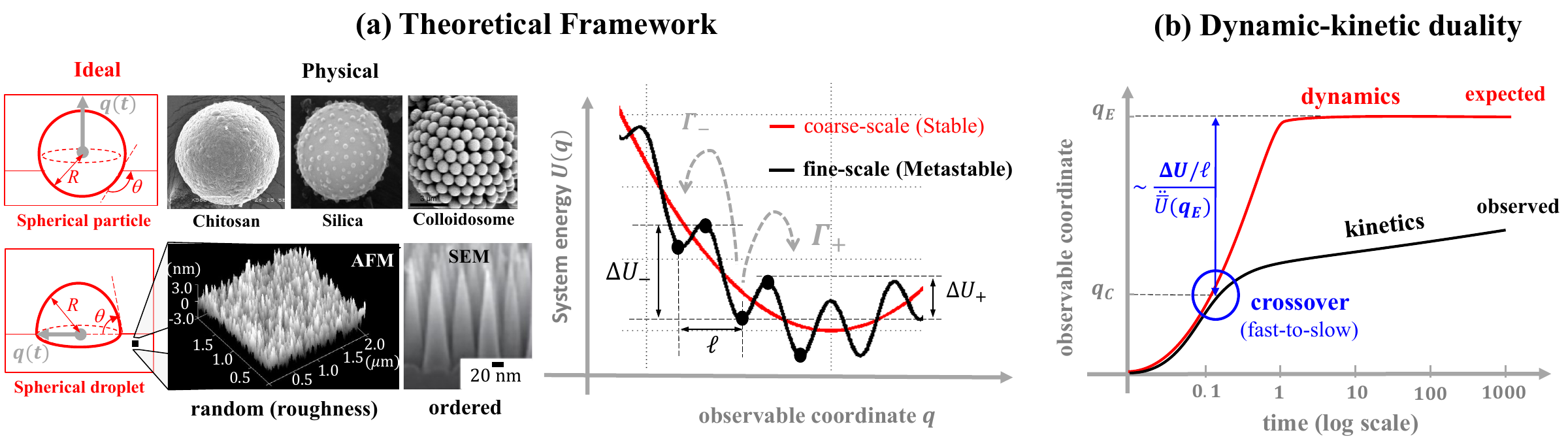}
\end{center}
\caption{{Dynamic-kinetic duality in particulate and multiphase systems.
(a) Theoretical framework: Coarse-scale idealizations neglect the micro/nanoscale surface structure present in physical systems. The nanoscale interfacial structure leads to energy landscapes with fine-scale metastability with period $\ell$ and characteristic magnitude $\Delta U$.
(b) Dynamic-to-kinetic crossovers, from ``fast'' dynamic evolution to ``slow'' thermally activated kinetics are observed beyond a crossover point $q_C$ as the system becomes critically close to the expected coarse-scale equilibrium at $q_E$.
}}
\label{fig:1}
\end{figure}

As a system with interfaces approaches the expected global equilibrium, the presence of nanoscale interfacial structure of physical or chemical nature (Fig.~\ref{fig:1}a) leads to a multiplicity of fine-scale metastable states, each satisfying \( \mathrm{d}U(\mathbf{q})/\mathrm{d}\mathbf{q} = 0 \), as illustrated schematically in Fig.~\ref{fig:1}b.
A key consequence of such fine-scale energy perturbations is that the observable system evolution departs drastically from the regimes where dynamic equations with well-defined damping coefficients apply. Instead, the system transitions to a regime governed by slower kinetics, described by rate equations with forward and backward rates \( \Gamma_\pm \) that are determinable using classical transition-state theory.\cite{kramers1940,hanggi1990,hanggi1986} In this regime, effective diffusivity and transport coefficients for describing the coarse-scale evolution become ``anomalous'' and are strongly dependent on the fine-scale structure of the energy landscape and the separation from equilibrium.\cite{colosqui2019diffusion}
Crossovers from ``fast'' non-equilibrium regimes governed by dynamic equations to ``slow'' or anomalous dynamic behaviors (Fig.~\ref{fig:1}b) at critical separations from coarse-scale equilibrium have been theoretically predicted and extensively documented by our group and coworkers for the case of micro- and nanoparticles at liquid-fluid and liquid-solid interfaces,\cite{colosqui2013,razavi2014,rahmani2016,keal2017,srivastava2020dual,colosqui2024kinetic,singletary2024surface,arnot2024electrochemistry} and, conversely, the displacement of liquid-fluid interfaces over surfaces with micro- or nanoscale topography.\cite{colosqui2015,colosqui2016,jose2018physical,colosqui2019diffusion,nandyala2020design,wang2021glass,zhao2023anomalous}
These studies identified the specific physical conditions and geometric parameters under which, for example, individual nano- or microparticles adsorb at different liquid interfaces, or single nano- or microdroplets spread on various nanostructured surfaces at controllable rates. These behaviors are quantitatively predicted by accounting for the fine-scale or ``glassy'' metastability induced by numerous nanoscale interfacial features.
The coarse-scale glassy or ``physical aging'' behavior due to a dense metastable landscape arising from the interplay between thermal motion, interface formation energy, and nanoscale topography can be effectively captured using classical sharp-interface continuum models, provided that the fine-scale energy perturbations introduced by nanoscale interfacial features are appropriately incorporated.

Accounting for fine-scale energy perturbations in the observable (coarse-scale) system evolution raises key challenges and opportunities, including how to promote or suppress the crossover from ``fast'' dynamic to ``slow'' kinetic regimes by tuning the energy barriers \( \Delta U \) and anomalously small spatial period \( \ell = A_d/L \) induced by interfacial nanostructure, or surface ``defects'', with characteristic area \( A_d \) of nanoscale dimensions, relative to the characteristic system dimension \( L \) (e.g., the particle, droplet, or pore radius).
An exciting opportunity from a better understanding of nanoscale interfacial structure is the translation of thermal fluctuations into directed transport via breaking the symmetry between forward and backward transition rates.\cite{bier1997brownian,kula1998brownian,hanggi2007brownian}
This perspective discusses the scientific foundations and experimental evidence and outlines the technological potential of controlling the dynamics and kinetic duality of particulate and multiphase systems. By reviewing how fine-scale metastability and thermal fluctuations influence coarse-scale transport processes, we highlight potential strategies to manipulate nanoparticle transport, liquid imbibition behavior, and the motion of fluid interfaces in both natural and engineered multiphase systems.

\section{Theoretical Framework~\label{sec:framework}}
This section summarizes the theoretical description and mathematical models enabling the rationalization of regime crossovers to ``anomalous'' regimes with unexpectedly slow relaxation to equilibrium that have been documented for various multiphase systems, particularly in the adsorption of colloidal particles at interfaces and in the imbibition of liquids within micro- and nanoscale confinements, which are discussed in the following section.

\subsection{Coarse-Scale Thermodynamic Potentials}
Sharp-interface thermodynamics is the building block for the description of particulate and multiphase systems. In this framework, we treat the interfacial region between two phases as a sharp, smooth dividing surface following Gibbs' original thermodynamic approach.\cite{gibbs1906scientific} The fundamental relations here can also be derived from more detailed thermodynamic descriptions that consider the interface as a thin but finite region where properties change gradually and thus have their own entropy and chemical potential.\cite{rowlinsonmolecular}
For simplicity we consider here the common case of an open thermodynamic system of volume $V$ consisting of three homogeneous phases ($i=1,2,3$) in thermodynamic equilibrium, separated by sharp interfaces with surface areas $A_{ij}$. Each phase contains $N_i^{(k)}$ molecules of substance $k$ ($k=1,2,\dots,N_s$) that can exchange mass with a large heat reservoir at temperature $T$.
At equilibrium, all phases have the same temperature $T_i = T$ and chemical potentials $\mu_i^{(k)} = \mu^{(k)}$, but not necessarily the same mechanical pressures $p_i$ since mechanical equilibrium involves interfacial forces. The idealization assumes interfaces occupy no volume, so $dV = \sum_i dV_i$. Increasing the interface area requires energy per unit area $\gamma_{ij}$, called the surface tension. The coarse-scale (i.e., observable scale) system energy is
\begin{equation}
\bar{U}(q) = T\,S - \sum_i p_i\,V_i + \sum_{i<j} \gamma_{ij}\,A_{ij} + \sum_{i,k} \mu^{(k)}\,N_i^{(k)}.
\label{eq:U}
\end{equation}
Summations run over phases $i$, substances $k$, and the three interfaces $A_{ij}$. For constant volume and temperature ($dV=0, dT=0$), the Helmholtz free energy $F=U-TS$ changes over coarse scales according to
\begin{equation}
\bar{F}(q) = - \sum_i p_i\,V_i + \sum_{i<j} \gamma_{ij}\, A_{ij} + \sum_{i,k} \mu^{(k)}\, N_i^{(k)}.
\label{eq:F}
\end{equation}
At constant temperature, volume, and chemical potentials ($dV=0, dT=0, d\mu^{(k)}=0$), it is useful to employ the coarse-scale grand potential
\begin{equation}
\bar{\Omega}(q)= - \sum_i p_i\,V_i + \sum_{i<j} \gamma_{ij}\, A_{ij}.
\label{eq:Omega}
\end{equation}
Changes in Helmholtz free energy represent reversible work in closed systems, while changes in grand potential represent reversible work in open systems. Depending on the studied conditions, it is convenient to use $\bar{U}$, $\bar{F}$, or $\bar{\Omega}$ to describe the coarse-scale system dynamics sufficently far from equilibrium.\\

\noindent{\bf Line Tension.} The expressions above neglect energy contributions \( dU_L = \tau dL \) caused by changes in the length \( L \) of the three-phase contact line, proportional to the line tension \( \tau \).\cite{widom1995line,marmur1997line} The line tension is the one-dimensional analog of surface tension, representing excess energy at the contact line where molecules interact among three phases. For typical molecular diameters \( \sigma \sim 0.1\, \mathrm{nm} \), line tension magnitude is about \( |\tau| \approx k_B T / \sigma \sim 10^{-11}\, \mathrm{N} \). Detailed theory estimates \( |\tau| \) between \( 10^{-12} \) and \( 10^{-10}\, \mathrm{N} \) for simple liquids, while experiments report values ranging from \( 10^{-8} \) to \( 10^{-6}\, \mathrm{N} \).\cite{amirfazli2004,erbil2014debate}
For simple molecular fluids with surface tension \( \gamma \sim 10^{-2} \mathrm{N/m} \), line tension effects become relevant for size scales \( \ell < \tau / \gamma \sim 1 \) to \( 10\, \mathrm{nm} \). Determining the effects of line tension for specific liquid-solid pairs and geometric configurations remains difficult and debated.\cite{kaz2011,erbil2014debate,huang2022effects}

\subsection{Fine-Scale Metastability}
\begin{figure}[htbp]
\centering
\begin{center}
\includegraphics[width=0.8\textwidth]{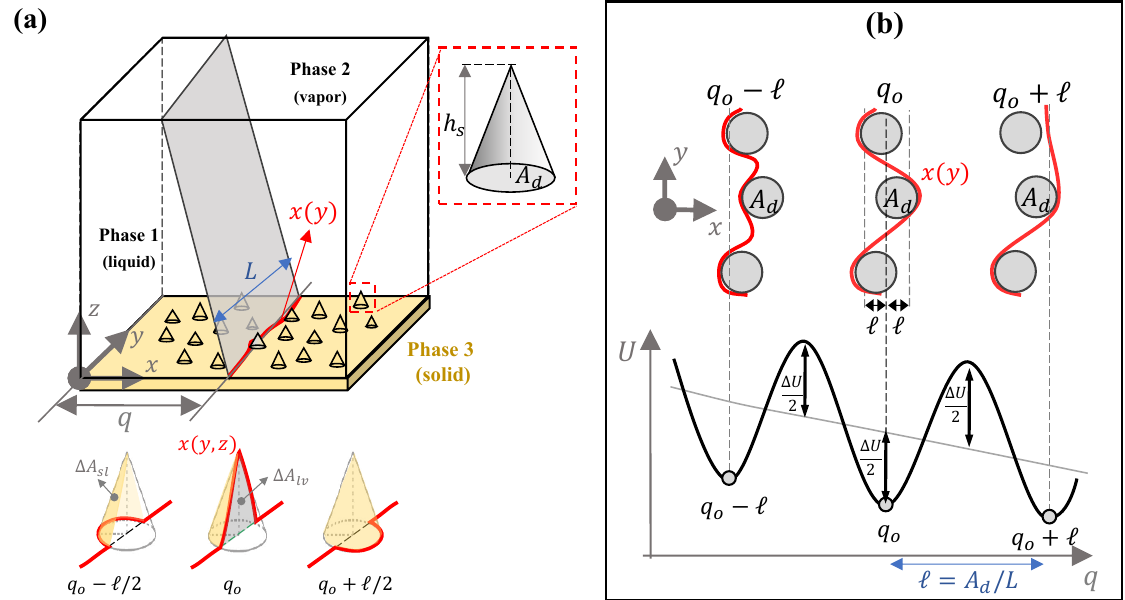}
\end{center}
\caption{Nanoscale surface topography and fine-scale metastability. (a) Wetting front of average perimeter $L$ at mean position $q$ along the $x$-direction, traversing a surface with numerous nanoscale topographic features of mean projected area $A_d$. Three-dimensional features are modeled as conical protrusions that induce highly localized variations of the interfacial surface, resulting in small surface area changes 
$\Delta A_{sl} \sim A_d$ and $\Delta A_{lv}\sim h_s\sqrt{A_d}$.
(b) Mean wetting front displacement $q=\bar{x}$ and energy profile $U(q)$ showing the emergence of energy barriers and numerous local (metastable) minima spaced by a fine-scale period $\ell=A_d/L_x\ll L$. \label{fig:2}
}
\end{figure}

We consider here the phenomenon of fine-scale metastability, focusing on the prototypical case of a liquid-fluid interface of total perimeter $L$ moving over a solid surface with nanoscale topographic features, as illustrated in Fig,~\ref{fig:2}. As is customary, the system energy fluctuations are parameterized by the (coarse-scale) observable $q = \bar{x}$, defined here as the average interface position $x(y)$ along the one-dimensional displacement direction for illustration purposes (Fig.~\ref{fig:2}a).
As illustrated in Fig.~\ref{fig:2}a, we consider an open system of fixed length, width, and height, where a sharp interface between two fluids moves along the $x$-direction over a flat solid surface densely populated by small defects with mean projected area $A_d$ and height $h_s$. For physical topographic features, their shape and characteristic dimensions can be obtained from topographic images with nanoscale resolution by Atomic Force Microscopy (AFM), Scanning Electron Microscopy (SEM), or by combining multiple techniques.\cite{checco2006,gujrati2018combining,aktar2020,chadha2024evaluating}

To construct simple expressions for inclusion in the thermodynamic potentials in Eqs.~\ref{eq:U}--\ref{eq:Omega}, we estimate the changes in interfacial surface areas $A_{sl}$ and $A_{lv}$, from changes in the local position $x(y,z)$ of the three-dimensional contact line perimeter over a single defect along the direction of motion Fig.~\ref{fig:2}a].
As illustrated in Fig.~\ref{fig:2}b, as the contact line moves over a single nanoscale defect, its average position $q$ increases by a small amount $\ell=A_d/L$.
Additionally, displacement of the contact line over a three-dimensional defect can cause an energy perturbation of magnitude \( \Delta U \) that we can estimate from the changes in surface area andthe corresponding surface energies. For systems of macro- or microscale dimensions (order 1 to 10 micrometers) with nanoscale defects (1 to 10 nanometers), the periods of energy perturbations are extremely small (0.1 to 1000 picometers). Because of the extremely large number of nanoscale features interacting with the much larger moving front or contact line, spatial variations in the thermodynamic potential can be effectively modeled by a single-mode perturbation
\begin{equation}
U(q) = \bar{U}(q) + \frac{1}{2} \Delta U 
\sin \left( \frac{2\pi q}{\ell} + \phi \right),
\label{eq:Utot}
\end{equation}
where \( \phi \) is an arbitrary phase shifting the global minimum by a small amount; hereafter we take \( \phi=0 \) for simplicity.\\

\noindent{\bf Energy Barriers from Nanoscale Surface Features}
The effective energy barrier $\Delta U$ in Eq.~\ref{eq:Utot} must account for two factors: (1) nanoscale chemical heterogeneities that cause small changes in local surface energies $\gamma_{sv}$ and $\gamma_{sl}$, and/or (2) nanoscale physical features that induce small changes in interfacial areas $A_{sv}$ and $A_{sl}$.
For a chemical heterogeneity on a flat surface changing the local Young contact angle by a small amount $\delta \theta$, one can estimate the energy barrier as
\begin{equation}
\Delta U \approx \gamma \lambda^2 \cos\theta \, \delta \theta,
\end{equation}
where $\gamma$ is the liquid-vapor surface tension.
For physical features such as hemispherical bumps, cylinders, or cones, the energy fluctuation can be estimated by geometric arguments. For example, for a cone of base area $A_d$ and height $h_s$ (see Fig.~\ref{fig:2}a), considering small changes $\delta A_{sv}$ and $\delta A_{sl}$ in interfacial areas as the contact line moves over the defect:

\begin{equation}
\Delta U \approx \gamma_{sv} \delta A_{sv} - \gamma_{sl} \delta A_{sl}.
\label{eq:DU}
\end{equation}
These expressions can provide reasonable estimates for mean energy barriers $\Delta U$ modulating the kinetic rates. Given the complexity of the multiscale nanoscale topography and chemical heterogeneities, the value of $\Delta U$ can be alternatively treated as a free parameter determined by fitting experimental results and the expressions above are employed to rationalize the employed quantities in terms of the mean defect area and r.m.s. topographic height.

\section{Dynamic-Kinetic Duality \label{sec:preliminary}}
A key prediction of the theoretical framework presented in Section~\ref{sec:framework} is that a crossover from dynamic regimes, governed by conventional transport equations parameterized by diffusion or damping coefficients, to anomalously slow relaxation to equilibrium, described by kinetic rate equations, is expected sufficiently close to the equilibrium state predicted for the state vector $q_E$ of coarse-scale (observable) variables. 
Based on a knowledge of the coarse-scale energy profile $\bar{U}(q)$ (Eqs.~\ref{eq:U}--Eqs.~\ref{eq:Omega}) one can predict anomalously slow kinetics for critically small separations from equilibrium
\begin{equation}
|q-q_E| \le \frac{\pi \Delta U }{\ddot{\bar{U}}_E\ell}
\label{eq:q-qE}
\end{equation}
where $\ddot{\bar{U}}_E\equiv d^2 \bar{U}(q_E)/d q^2$ is the coarse-scale curvature of the energy profile at equilibrium and 
\begin{equation}
\ell = A_d/L
\label{eq:ell}
\end{equation}
is the fine-scale perturbation period, which is determined by the characteristic projected area $A_d$ of the nanoscale surface features and the characteristic length or perimeter $L$ of the moving weeting front or contact line. For example, $L=2\pi R$ for a spherical particle or circular pore of radius $R$.

It is worth noting that the perturbation period along the $q$-coordinate defined by Eq.~\ref{eq:ell} can be extremely small since it is the ratio of a nanoscale surface area and the macro- or microscale characteristic length of the contact line or wetting front.
For example, $\ell\sim{\cal O}(10^{-12} m)$ for the case of 1-$\mu$m colloidal particle with 1-nm topographic defects.\cite{kaz2011,colosqui2013,keal2017}\\

\noindent{\bf The Kinetic Regime.\label{sec:kinetic}}
When a particulate or multiphase system is critically close to equilibrium, as predicted by Eq.~\ref{eq:q-qE}, the average evolution along the $q$-coordinate can be determined by the ``forward/backward'' (+/-) transition rates\cite{kramers1940,colosqui2013,colosqui2017} 
\begin{equation}
\Gamma_{\pm}
= \frac{1}{2\pi\xi}
\sqrt{\frac{\partial^2 U(q_o)}{\partial z^2} \left|\frac{\partial^2 U(q_\pm)}{\partial q^2}\right|} \exp\left(-\frac{\Delta U_{\pm}}{k_B T}\right),
\label{eq:kramers}
\end{equation}
where $\Delta U_{\pm}=U(q_\pm)-U(q_o)$ are energy the barriers opposing ``forward/backward'' displacements between a local minimum $z_o$ and neighboring maxima $q_{\pm}=q_o\pm \ell/2$ (see Fig.~\ref{fig:2}b). 
The prefactor in Eq.~\ref{eq:kramers} is prescribed by the damping coefficient $\xi$, determined from energy dissipation, and the curvature of the energy profile at the local maxima and minima.
When the separation between neighboring minima is given by a period $\ell$, the time evolution of the noise-averaged evolution coordinate $\langle q(t) \rangle$ is descibed by a rate equation
\begin{equation}
\langle \dot{q} \rangle =\langle dq/dt \rangle=\ell (\Gamma_{+}-\Gamma_{+}).  
\label{eq:rate}
\end{equation}
\noindent According to Eqs.~(\ref{eq:kramers}--\ref{eq:rate}), displacement rates decrease exponentially with the energy barrier $\Delta U_\pm$ induced by surface features of nanoscale dimensions.
Notably, {single-digit nanoscale features (1-10~nm) can induce significant energy barriers $\Delta U\sim \gamma A_d \gtrsim 10 k_B T$} and extremely slow kinetic rates for the case of simple liquids with characteristic molecular diameters $\sigma\sim 0.3$~nm, which have surface energies $\gamma\sim k_BT/\sigma^2$; here $k_B T\sim{\cal O}(10^{-21} \mathrm{J})$ is the thermal energy.
Therefore, proper combinations of interfacial feature and macroscale system dimensions can drastically {advance or delay the crossover point} to kinetic regimes. 
%

Two key aspects of applying classical Kramers theory within the proposed framework leading to Eqs.~\ref{eq:kramers}--\ref{eq:rate} are often overlooked. 
First, the extremely small fine-scale perturbation period $\ell$ required in Eq.~\ref{eq:rate} to fit experimental data is set by ratios of physically meaningful, measurable dimensions.\cite{colosqui2013,rahmani2016,keal2017,colosqui2016,zhao2023anomalous} 
The need for anomalously small values, $\ell \le {\cal O}(10^{-9})$~m, much smaller than any physical feature dimension, is a well-documented outcome of molecular kinetic theory and related models.\cite{haynes1969,blake2011} 
Second, in the present formulation, the damping coefficient $\xi$ is prescribed by the actual dissipation of energy, rather than treated as an effective parameter with unrealistically large values to match observed contact-line or wetting-front kinematics~\cite{haynes1969,blake2011,duvivier2011experimental,duvivier2013,sedev2015molecular}. 
Accordingly, $\xi$ in the proposed framework satisfies mechanical and thermal energy conservation and can be obtained from conventional fluid-mechanical models~\cite{colosqui2013,rahmani2015,colosqui2015,colosqui2016} or from fluctuation-dissipation relations.\cite{rahmani2016,boniello2015brownian}

\subsection{Micro/Nanoparticles at Liquid Interfaces}
A seminal study by Manoharan and co-workers,\cite{kaz2011} provided experimental evidence and a theoretical framework showing that the adsorption of microscale particles at liquid-liquid interfaces can be far slower than predicted by conventional dynamic models based solely on capillary forces and viscous drag.\cite{kaz2011} This phenomenon, termed \emph{physical ageing} and analogous to slow relaxation in glassy systems, was linked to thermally activated hopping of the contact line over nanoscale surface heterogeneities or nanoscale roughness.
Subsequent studies explored its implications for emulsion stabilization, interfacial self-assembly, and particle-laden interface dynamics.\cite{garbin2012,schwenke2014assembly,sharifi2016capillary,wang2016contact,hu2020particle}

\begin{figure}[h!]
\begin{center}
\includegraphics[width=0.97\textwidth]{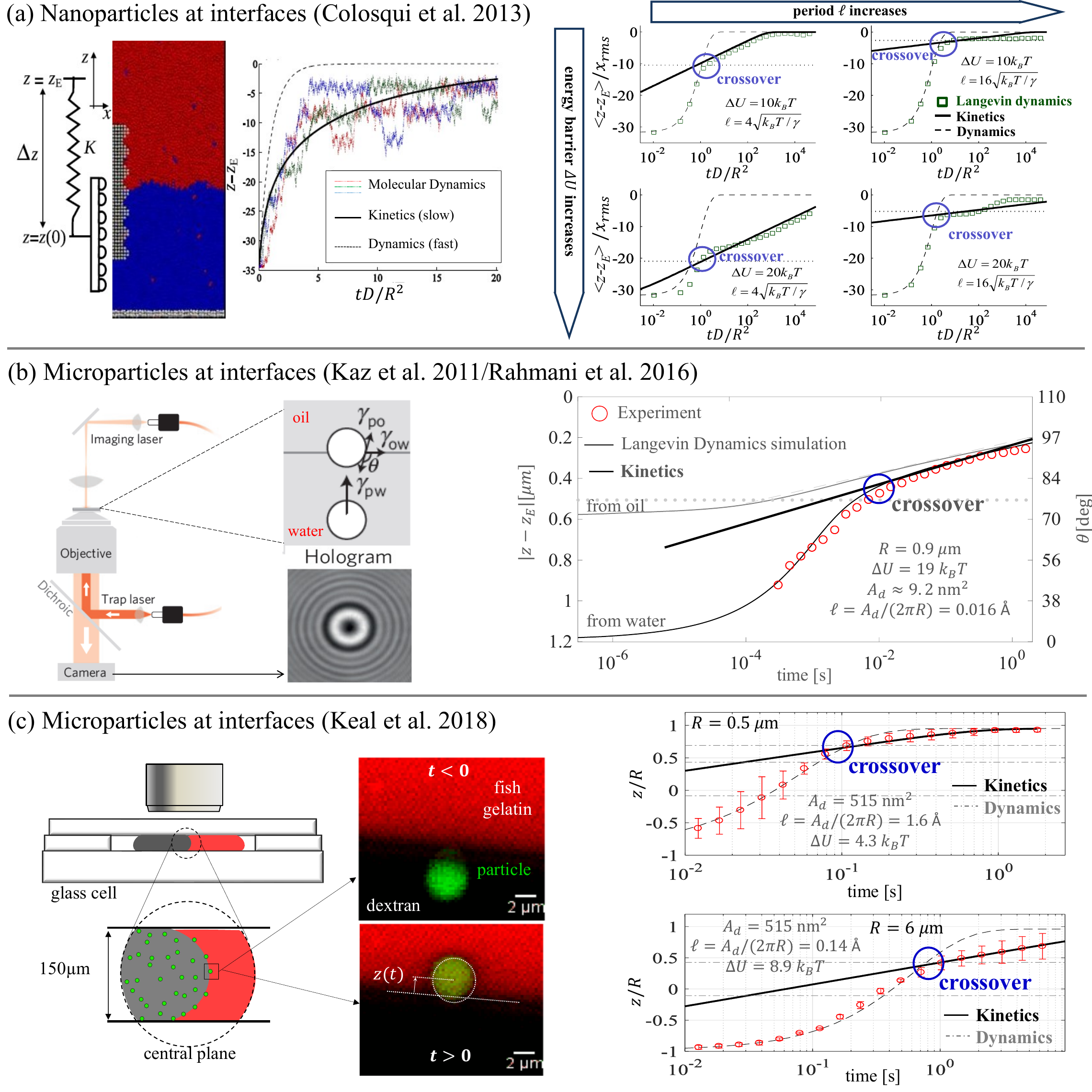}
\end{center}
\caption{Particles at liquid interfaces. The observable coordinate $q=z$ is the normal distance from the liquid interface, the coarse-scale equilibrium is expected at $z=z_E$.  
(a) Molecular dynamics simulations of nanoparticles with small surface features reveal fine-scale metastability and dynamic-kinetic regime crossover as $|z-z_E|$ decreases. Langevin dynamics simulations verify predictions by Eq.~\ref{eq:q-qE} for controlling the crossover point through the energy barrier $\Delta U$ and period $\ell$ [adapted from Ref.~\citenum{colosqui2013}].
(b) Experimental results for microparticles $R=0.9$~$\mu$m at oil-water interfaces from Kaz et al. 2011 \cite{kaz2011} are accounted for by Langevin simulations and the rate equation (Eq.~\ref{eq:rate}) for equilibrium distance $|z-z_E|$ using physically justified parameters.\cite{rahmani2016}.
(c) Microparticle adsortion at water/water interfaces report dynamic-to-kinetic regimes crossovers predicted by Eq.~\ref{eq:q-qE} for different particle radius $R=$~0.5 \& 6~$\mu$m [adapted from Ref.~\citenum{keal2017}].
\label{fig:3}}
\end{figure}

Key fundamental aspects and implications of the anomalously slow relaxation phenomenon reported by Manoharan and co-workers were further elucidated using the theoretical framework described in Sections 2 and 3, and were verified by molecular dynamics (MD) and Langevin simulations of nanoparticles at liquid interfaces (Fig.~\ref{fig:3}a).\cite{colosqui2013,razavi2014,rahmani2015,rahmani2016}
An important finding of MD simulations\cite{colosqui2013} is that three-dimensional localized pinning of the interface between two fluid phases at nanoscale surface features smaller than 1 nm leads to the appearance of long-lived metastable states and a predictable transition to a much slower relaxation to equilibrium as the system approaches the known equilibrium state, prescribed by forcing the system (see Fig.~\ref{fig:3}a).
Langevin simulations further reported that modifying the energy barrier $\Delta U$, by increasing the feature size and/or surface tension, and the period $\ell$ can drastically advance or delay the crossover point to the anomalously slow kinetic regime, as shown in Fig.~\ref{fig:3}a.
Further, Langevin dynamics simulations by Rahmani et al. (2016)\cite{rahmani2016} (Fig.~\ref{fig:3}b) were able to account for the experimental observations for microparticles ($R=0.9$~$\mu$m) in water-oil interfaces by Kaz et al. (2011), by incorporating the fine-scale energy fluctuations induced by nanoscale surface defects via Eq.~\ref{eq:Utot}.
As shown in Fig.~\ref{fig:3}b, these mesoscale simulations, which resolve the fine-scale energy landscape, effectively capture the entire evolution process, including both dynamic and kinetic regimes. Using the nanoscale defect area $A_d \simeq 9.2\,\text{nm}^2$ as an input parameter physically justifies the emergence of extremely small periods $\ell = \frac{A_d}{2\pi R} \simeq 1.63\,\text{pm}$ between the fine-scale (unobservable) metastable states that give rise to the observable coarse-scale evolution.\cite{rahmani2016}

Further experimental work using confocal fluorescence microscopy\cite{keal2017} (see Fig.~\ref{fig:3}c) corroborated the ability to predict the crossover from dynamic to kinetic regimes using Eq.~\ref{eq:q-qE}, and the subsequent slow relaxation to equilibrium using Eqs.~\ref{eq:kramers}--\ref{eq:rate} for the case of latex microparticles of different radius \(R \approx 0.5-6\,\mu\text{m}\) in water-in-water emulsions with extremely low surface tension (Fig.~\ref{fig:3}c).
In this study, energy barriers \(\Delta U \simeq 4-9\,k_B T\) and periods \(\ell = A_d/(2 \pi R) \simeq 0.1 - 1\, \text{nm}\), used in Eqs.~\ref{eq:kramers}--\ref{eq:rate}, were determined by using a single feature area \(A_d \simeq 500\,\text{nm}^2\) for all the particle dimensions, which was attributed to the adhesion of polymer chains on the particle surface.
The experimental work by Keal et al.\cite{keal2016,keal2017} provided solid evidence that the crossover from fast dynamic decay to anomalously slow relaxation to equilibrium is controlled by the ratio of surface feature area to particle radius (cf. Fig.~\ref{fig:3}c).

\subsection{Liquid Interfaces on Micro/Nanostructured Surfaces}
\begin{figure}[h!]
\begin{center}
\includegraphics[width=0.95\textwidth]{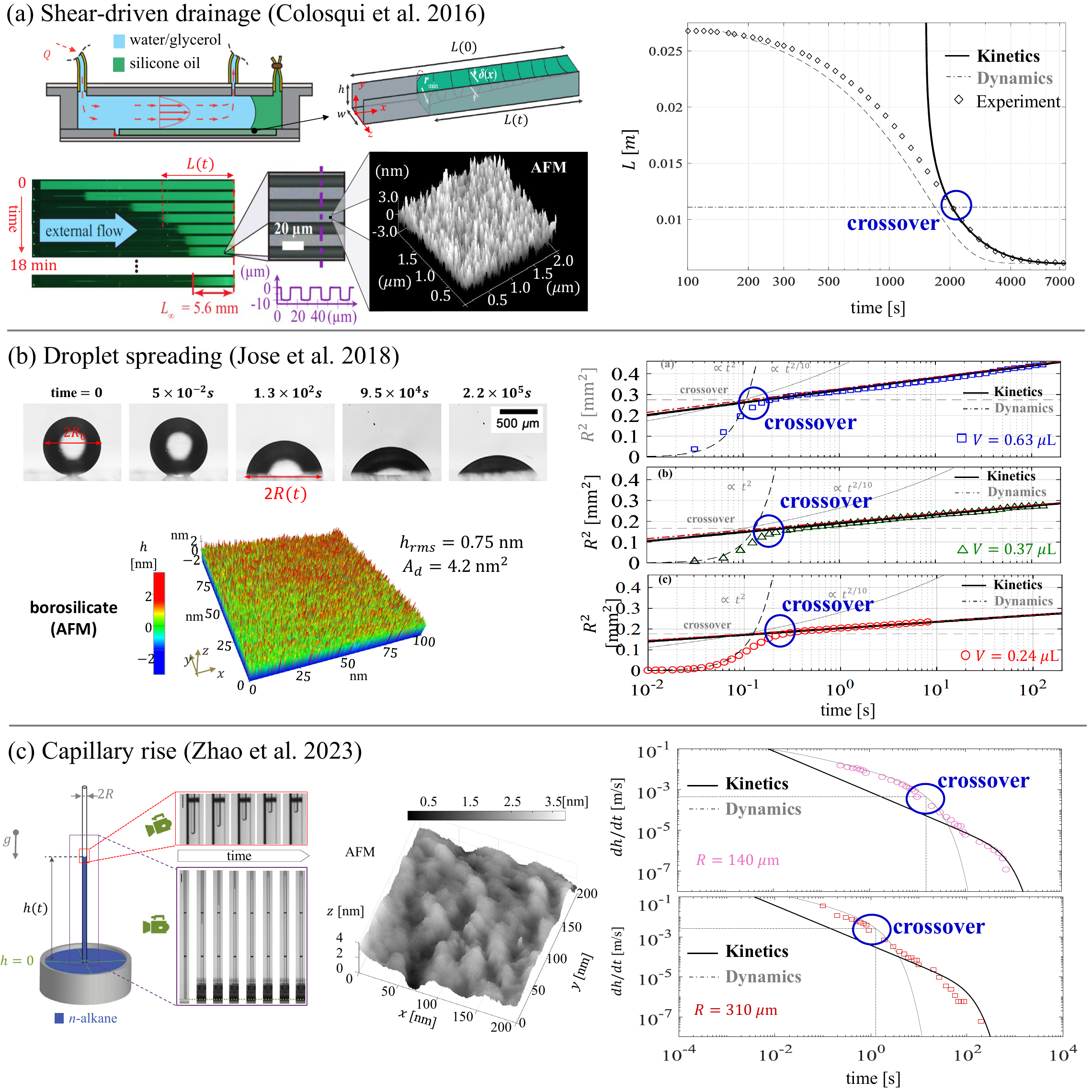}
\end{center}
\vskip -15 pt
\caption{Liquid interfaces on surfaces with micro/nanoscale topography. In all cases the crossover point is predicted by Eq.~\ref{eq:q-qE} for the corresponding observable coordinates, and the kinetic regime evolution is given by Eqs.~\ref{eq:kramers}--\ref{eq:rate}. 
(a) Shear-driven drainage of square microcapillaries report dynamic-kinetic transition for the retained liquid length $L(t)$, using a defect area $A_d$ determined by AFM imaging [Adapted from Ref.~\citenum{colosqui2016}] 
(b) Droplet spreading (water in oils) by Jose et al. (2011)\cite{jose2018physical} report a slow (over 1-hour) logarithmic spreading regime for the contact area $q=R^2$ emerging after initial power-law spreading. Observations for different droplet volume are accounted for using a single surface defect $A_d$ determined by AFM [images from Ref.~\citenum{jose2018physical}].   
(c) Anomalous capillary rise of alkane oils report a remarkable slow approach to equilibrium, with the rise rate $dh/dt$ showing dynamic-kinetic crossover at column heights $h$ predicted by Eq.~\ref{eq:q-qE}. Observations for different capillary radius ($R=$~140 \& 310~$\mu$m) are predicted by a single defect area $A_d$ estimated via AFM analysis [adapted from Ref.~\citenum{zhao2023anomalous}].
\label{fig:4}}
\end{figure}

The theoretical models presented in Section 2 describe bot nanoparticles at interfaces and liquid interfaces moving over surfaces with micro- or nanoscale features, by capturing fine-scale metastability arising from energy fluctuations due to nanoscale interfacial structure. As the interface advances, these physical and chemical heterogeneities create a ``rugged'' energy landscape with numerous metastable states, causing highly localized pinning of the wetting front and thermally activated hopping between states. Hence, by incorporating fine-scale energy fluctuations, the proposed models explain transitions from ``fast'' capillarity-driven spreading and imbibition to slow, kinetically limited regimes reported for common wetting and imbibition processes.

Analytical predictions from Eq.~\ref{eq:kramers}--\ref{eq:rate} have accounted for experimental observations of capillary imbibition in microcapillaries and spreading of microdroplets (see Fig.~\ref{fig:4}).
Work by Wexler et al. studying the shear-driven drainage and failure of liquid-infused surfaces\cite{wexler2015} found a finite liquid column length \( L_E \) retained under equilibrium of capillary and shear forces.
Careful analysis of the liquid column evolution revealed a dramatic crossover from conventional Lucas-Washburn dynamics for the observable column length \( q = L(t) \) to an anomalously slow relaxation regime when the column was smaller than critical values predicted by Eq.~\ref{eq:q-qE} (Fig.~\ref{fig:4}a).\cite{colosqui2016}   
The forced drainage of the liquid near the expected equilibrium followed the slow evolution predicted by Eqs.~\ref{eq:kramers}--\ref{eq:rate} for various applied shear rates using a single characteristic feature area \( A_d \simeq 0.004\,\mu\mathrm{m}^2 \) that was determined by topographic atomic force analysis (Fig.~\ref{fig:4}a).
The defect area employed to account for experimental observations corresponded to nanoscale topographic features with average lateral dimension \( s_d \simeq 35\,\mathrm{nm} \) and height \( h_d \simeq 1\,\mathrm{nm} \), which produced a characteristic energy barrier \( \Delta U = 3.4\,k_B T \) predicted via Eq.~\ref{eq:DU} and a period \( \ell = 0.15\,\mathrm{nm} \) determined as the ratio of the defect area \( A_d \) to the microcapillary perimeter.\cite{colosqui2016} 

The transition from dynamic to kinetic regimes was additionally reported in droplet spreading experiments performed by Jose et al.\cite{jose2018physical} (Fig.~\ref{fig:4}b).
This study showed that the wetting radius evolution \( R(t) \) for a millimeter-sized water droplet immersed in viscous silicon oil (\( \mu_{\mathrm{oil}} = 100 \) cSt) that spreads on a borosilicate substrate can take several hours to reach equilibrium (Fig.~\ref{fig:4}b), similarly to the physical ageing of microparticles first reported by Manoharan et al. in 2011.
The crossover to the anomalously slow spreading regime, which is nearly logarithmic in time (\( R \propto \log t \)) (Fig.~\ref{fig:4}b), emerges after the characteristic power-law spreading dynamics (\( R \propto t^n \)) at short initial times.\cite{bird2008short,eddi2013short}
As in previous studies, the regime crossover and slow relaxation rates were predicted using a single energy barrier \( \Delta U \simeq 11-17\,k_B T \) determined from a characteristic surface feature \( A_d = 4.2\,\mathrm{nm}^2 \) estimated from AFM topographic analysis.\cite{jose2018physical}  

The most recent experimental evidence, and arguably the most compelling case for fine-scale metastability inducing dynamic-kinetic regime crossovers, is provided in the experimental study by Zhao et al.\cite{zhao2023anomalous} on the spontaneous imbibition of simple alkane oils in glass microcapillaries.
This work showed that the common phenomenon of capillary rise in capillary tubes exhibits a dramatic transition from well-known Lucas-Washburn dynamics to an anomalously slow regime predicted by the kinetic model in Eq.~\ref{eq:kramers}--\ref{eq:rate} (Fig.~\ref{fig:4}c).
In this case, the state variable or observable \( q = h(t) \) is the liquid column height above the reference level, and the experimental study by Zhao et al. followed the evolution for over 10 hours without attaining the expected equilibrium in some conditions.
The column height at which the dynamic-to-kinetic crossover occurs (Eq.~\ref{eq:q-qE}) and the subsequent relaxation (Eq.~\ref{eq:rate}) were accounted for by energy barriers \( \Delta U \simeq \) and a single surface defect area \( A_d = \) for four different oils and two different capillary tube diameters (Fig.~\ref{fig:4}c).

The studies presented in Fig.~\ref{fig:4} provide a volume of experimental evidence highlighting how the interplay between capillary geometry and nanoscale surface topography critically influences the crossover from fast dynamic wetting to much slower, thermally activated kinetic regimes, that should not be misinterpreted as dynamic conditions with anomalously high damping coefficients and high dissipation of energy.

\section{From Fine-Scale Metastability to Coarse-Scale Transport\label{sec:ratchets}} 
\begin{figure}[h]
\begin{center}
\includegraphics[width=\textwidth]{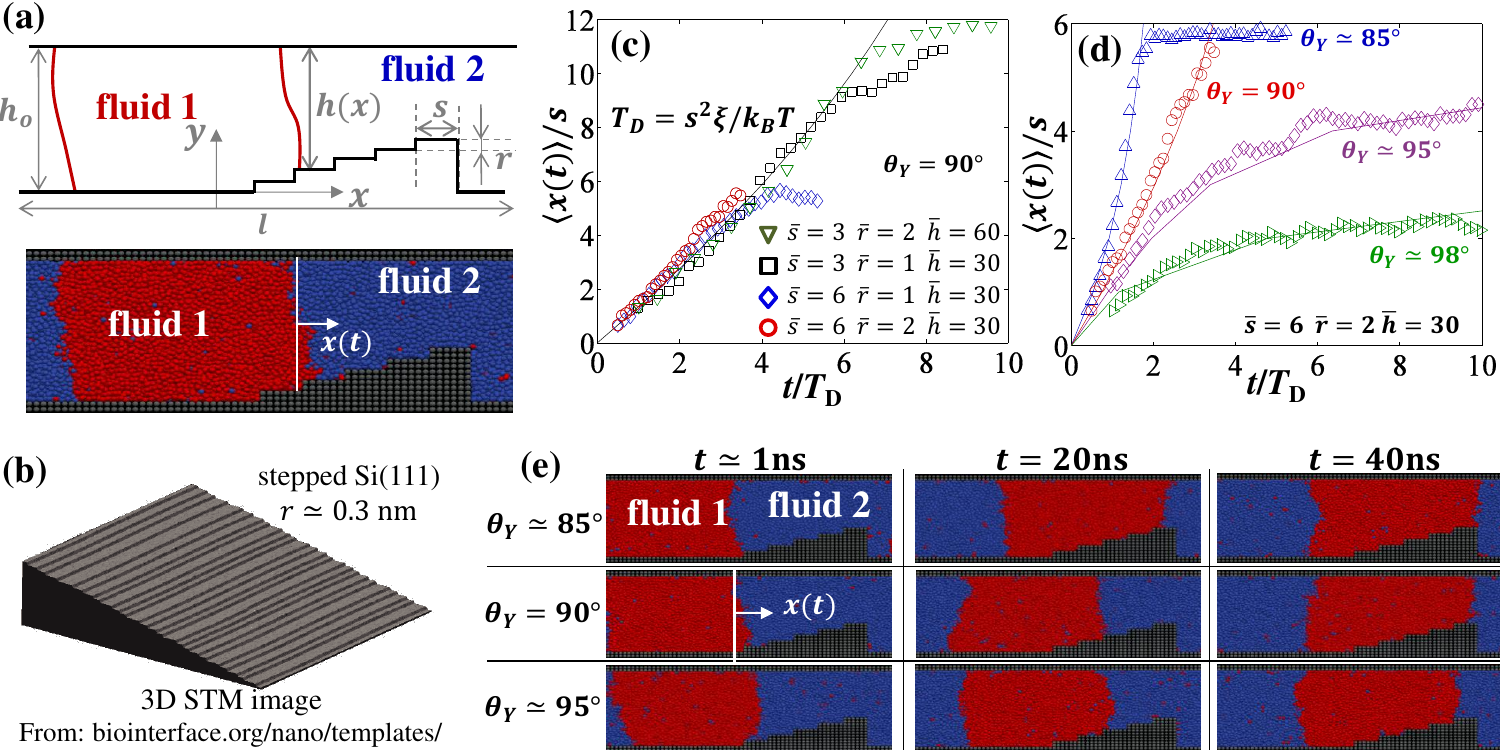}
\end{center}
\caption{Rectification of thermal motion into directed liquid transport in terraced nanopores [adapted from Ref.~\citenum{colosqui2015}]. 
(a) Slit nanopore with a terraced surface (step \( s \), riser \( r \)).
(b) Atomically stepped Si(111) (\( r \sim 0.3\,\text{nm} \)) [3D STM image from biointerface.org]. 
(c--d) Mean displacement \(\langle x(t) \rangle\) for different terraced geometry (\(\overline{s} = s/\Delta x\), \(\overline{r} = r/\Delta x\), \(\overline{h} = h_0/\Delta x\)) and wettability conditions (\(85^\circ \lesssim \theta_Y \lesssim 100^\circ\)).  
Lines: analytical solution via Eqs.~\ref{eq:smoluchowski}--\ref{eq:meanx}. Markers: MD simulation (\(T_D = s^2 \xi / k_B T\), \(\Delta x \simeq 0.3\,\text{nm}\)).
(e) MD simulation snapshots for different liquid-solid affinity with \(\theta_Y = 85^\circ, 90^\circ, 95^\circ\) at time instances \(t \simeq 1, 20, 40\) ns.  
\label{fig:5}}
\end{figure}
%
When ``forward'' and ``backward'' transition rates between metastable states are asymmetric (\(\Gamma_+ \neq \Gamma_-\)), as in the cases of terraced surface topography of nanoscale step width $s$ and height $r$ (see Fig.~\ref{fig:5}a), thermal fluctuations can rapidly drive the system toward the direction where energy barriers are lower. 
Asymmetric energy barriers (\(\Delta U_+ \neq \Delta U_-\)) separating metastable states can thus rectify thermal motion into directed transport, as in the case of Brownian ratchets.\cite{bier1997brownian,kula1998brownian,hanggi2007brownian} 
Brownian ratchets underlie fundamental biological processes by which thermal motion is harnessed to perform directed transport and net work.\cite{simon1992drives,hille2001ion,nelson2004biological,hanggi2009artificial} Similarly, surface nanostructures with intentionally designed directional variations in local feature height and/or projected area can convert random thermal motion of interfaces into directed transport of liquids, solutes, electrolytes, and charge carriers.\cite{schoch2008transport,xiao2019ion,huang2020molecular,liu2024ion} 

To model the rectification of thermal motion in liquid imbibition, we begin by considering that the evolution of liquid infiltration in a pore is parameterized by a coarse-scale coordinate \( q = x \) given by the average position $x$ of all the atoms or molecules composing the liquid interface (see Fig.~\ref{fig:5}a)).
We can then project the system energy prescribed by the position of every atom onto a one-dimensional profile $U(x)$ with fine-scale metastable states (Eq.~\ref{eq:Utot}), parameterized by the (observable) average interface position $x(t)$.
The random thermal motion of the interface is statistically described by a probability density $p(x,t)$ governed by the one-dimensional Smoluchowski diffusion equation:
\begin{equation}
\frac{\partial}{\partial t} p(x,t) = \frac{\partial}{\partial x} \frac{1}{\xi(x)} \left[ k_B T \frac{\partial}{\partial x} + \frac{\partial U}{\partial x} \right] p(x,t),
\label{eq:smoluchowski}
\end{equation}
where $k_B T$ is the system thermal energy and $\xi(x)$ is the local damping coefficient, consistent with dissipation-dissipation relations so that the interface diffusivity is $D=k_B T/\xi$.
The noise-average or expected interface displacement is then given by
\begin{equation}
q=\langle x(t) \rangle = \int_{-\infty}^\infty x\, p(x,t) \, dx,
\label{eq:meanx}
\end{equation}
and can be obtained via analytical solutions for simple geometries or numerical simulation of Eq.~\ref{eq:smoluchowski}.\cite{colosqui2015}

Our group has proposed and applied the theoretical framework based on Eq.~\ref{eq:smoluchowski} and energy profile in Eq.~\ref{eq:Utot} to predict liquid transport and imbibition in nanostructured pores with nanoscale ``terraced'' surfaces of different dimensions (see Fig.~\ref{fig:5}a-b).\cite{colosqui2015} 
Theoretical predictions from Eqs.~\ref{eq:smoluchowski}--\ref{eq:meanx} and MD simulations indicate that properly designed nanostructures can induce very fast thermally activated infiltration of liquid into small nanopores (Fig.~\ref{fig:5}c), even when the pore material is hydro/oleophobic (i.e., the Young contact angle is \(\theta_Y > 90^\circ\)) and one would expect the spontaneous drainage of liquid instead (see Fig.~\ref{fig:5}f).
Notably, the liquid interface displacement rates induced by the studied nanostructures (cf. Fig.~\ref{fig:5}e) range between 0.1 and 1 m/s for a slit nanopore with nanoscale dimensions (\( l = 25\,\text{nm} \), \( h_0 = 10{-}20\,\text{nm} \)).

These MD study supports the theoretical prediction that physical surface nanostructures and properly engineered pore geometries can drastically enhance liquid transport, achieving flow rates anomalously larger than hydrodynamic predictions based solely on viscous shear in nanoscale confinement.
Moreover, the physical surface structure can enable the system to evolve away from the expected coarse-scale equilibrium, with the confined liquid moving against the direction predicted by conventional coarse-scale thermodynamic potentials for liquid imbibition.
These predictions verified by MD simulations call for experimental validation in future studies, strategically guided by the theoretical framework delineated in this article.  

\section{Technological Impact and Outlook}

The theoretical framework and volume of experimental observations for different multiphase systems discussed in this article highlight that precise prediction and control of the system evolution is enabled by quantifying the effects of fine-scale energy barriers induced by nanoscale surface defects on macroscopic transport properties. 
These insights can have profound implications for designing technical applications of particulate and multiphase systems for diverse microfluidic systems, surface coatings, membranes, and porous materials where tailoring dynamic and kinetic behaviors can enhance performance, stability, or selectivity. 

The dynamic-kinetic duality framework presented in this article opens a path from fundamental sharp-interface thermodynamics to transformative applications of particulate and multiphase systems in materials synthesis, separation processes, energy conversion and harvesting, using nanostructured surfaces with natural or engineered features. By understanding how nanoscale interfacial features govern fine-scale metastability, energy barriers, and rate asymmetries, for system of different dimensions, one can design and control the time scale of transport processes in particulate and multiphase systems, whether the goal is to accelerate particle adsorption, delay droplet spreading and imbibition, or rectify interfacial thermal motion into useful work.
Technical applications that can directly benefit from these insights span emulsion stabilization,\cite{binks2002solid,san2012influence,zanini2017} targeted drug delivery,\cite{tao2003micro,prabaharan2004,hughes2005nanostructure,ghosh2008gold} additive manufacturing,\cite{sachs1993three,gibson2010additive,n2014review} and colloidal assembly at interfaces.\cite{mcgorty2010,furst2011directing,oleg2014,srivastava2020dual} Controlling adsorption and desorption kinetics through energy landscape engineering enables programmable  colloidal assembly,\cite{mcgorty2010,juarez2012feedback,zhang2016forming} or controlling charge transfer by electroactive nanomaterials at wetted electrode surfaces.\cite{kissling2011electrochemical,hoyt2016modeling,tang2022theoretical,arnot2024electrochemistry}  
In emulsions, fine-scale metastability models can help predict long-term stability and tailored droplet properties,\cite{zanini2017,zanini2019mechanical} and in additive manufacturing, these models guide precise droplet deposition and spreading on structured surfaces.\cite{sachs1993three,gibson2010additive} Furthermore, the design of omniphobic surfaces\cite{dupuis2005,tuteja2007,bhushan2009a,dufour2010,paxson2013self,rykaczewski2014,nandyala2020design} leverage these models to engineer customized wettability and repellency.

Important opportunities for future modeling developments lie in integrating the predictive framework described in this article with 3D mesoscale simulation methods\cite{glotzer2002molecular,SM2012,PRL1,PRE2010,seaton2013dl_meso} for the coarse-scale evolution of particulate and multiphase systems and machine learning techniques for mulstiscale simulations to more efficiently model the (unresolved) fine-scale dynamics effects on (resolved) the coarse-scale evolution.\cite{lubbers2020modeling,ingolfsson2023machine} 
Basic machine learning techniques are suitable to infer the fine-scale structure of energy potentials that account for large number of direct observations of the macroscale evolution.\cite{lee2020coarse,karniadakis2021physics}
Furthermore, coupling nanoscale imaging and metrology with multiscale modeling can enable rapid mapping from surface geometry and chemistry to coarse-scale evolution and macroscopic properties of the system. In parallel, advances in nanofabrication, self-assembly, directed deposition, and hybrid lithographic methods, will expand the design space of interfacial architectures that achieve desired dynamic-kinetic crossovers and kinetic regimes on demand.

Ultimately, the more accurate theoretical description of fine-scale metastability, dynamics and kinetics, and regime crossovers not only interpret complex experimental phenomena but also provide a predictive foundation for the rational design of next-generation applications of particulate and multiphase systems that exploit nanoscale interfacial features to control transport behavior with unprecedented precision.


\end{document}